\documentclass[reprint,aps,prb,floatfix,superscriptaddress,showpacs]{revtex4-1}
\usepackage{mathtools}
\usepackage{amssymb}
\usepackage{graphicx}
\usepackage{hyperref}

\hypersetup{
  colorlinks   = true, 
  urlcolor     = blue, 
  linkcolor    = blue, 
  citecolor   = blue 
}

\begin{document}


\title{Controllable quantum scars in semiconductor quantum dots}

\author{J. Keski-Rahkonen}
\affiliation{Laboratory of Physics, Tampere University of Technology, Finland}
\author{P. J. J. Luukko}
\affiliation{Max Planck Institute for the Physics of Complex Systems, Dresden, Germany}
\author{L. Kaplan}
\affiliation{Department of Physics and Engineering Physics, Tulane University, New Orleans, USA}
\author{E. J. Heller}
\affiliation{Department of Physics, Harvard University, Cambridge, USA}
\author{E. R\"as\"anen}
\affiliation{Laboratory of Physics, Tampere University of Technology, Finland}
\affiliation{Department of Physics, Harvard University, Cambridge, USA}

\date{\today}


\begin{abstract}
  Quantum scars are enhancements of quantum probability density along classical
  periodic orbits. We study the recently discovered phenomenon of strong,
  perturbation-induced quantum scarring in the two-dimensional harmonic
  oscillator exposed to a homogeneous magnetic field. We demonstrate that both
  the geometry and the orientation of the scars are fully controllable with a
  magnetic field and a focused perturbative potential, respectively. These
  properties may open a path into an experimental scheme to manipulate electric
  currents in nanostructures fabricated in a two-dimensional electron gas.
\end{abstract}

\maketitle

\section{Introduction}

Quantum scars~\cite{Heller, Kaplan} are tracks of enhanced probability density
in the eigenstates of a quantum system. They occur along short unstable periodic
orbits (POs) of the corresponding chaotic classical system. While being an
interesting example of the correspondence between classical and quantum
mechanics, quantum scars also show how quantum mechanics may suppress classical
chaos and make certain systems more accessible for applications. In
particular, the enhanced probability density of the scars provides a path for
quantum transport across an otherwise chaotic system.

A new type of quantum scarring was recently discovered by some of the present
authors.~\cite{Luukko} It was found that local perturbations (potential
``bumps'') on a two-dimensional (2D), radially symmetric quantum well produce
high-energy eigenstates that contain scars of short POs of the corresponding
\emph{unperturbed} (without bumps) system. Even though similar in appearance to ordinary quantum scars,
these scars have a fundamentally different origin. They result from
resonances in the unperturbed classical system, which in turn create
semiclassical near-degeneracies in the unperturbed quantum system. Localized
perturbations then form scarred eigenstates from these near-degenerate
``resonant sets'' because the scarred states effectively extremize the
perturbation. These perturbation-induced (PI) scars are unusually strong
compared to ordinary scars, to the extent that wave packets can be transported
through the perturbed system, along the scars, with higher fidelity than
through the unperturbed system, even with \emph{randomly placed}
perturbations.~\cite{Luukko}

In this work we study these perturbation-induced scars in a 2D quantum harmonic
oscillator exposed to a perpendicular, homogeneous magnetic field. This system
has direct experimental relevance as it is a prototype model for semiconductor
quantum dots in the 2D electron gas.~\cite{reimannmanninen} Previous studies
combining experiments and theory have confirmed the validity of the harmonic
approximation to model the external confining potential of electrons in the
quantum dot, up to the quantum Hall regime reached with a strong magnetic
field.~\cite{rasanen,rogge} Furthermore, the role of external impurities in 2D
quantum dots has been quantitatively identified through the measured
differential magnetoconductance that displays the quantum
eigenstates.~\cite{konemann} We show that once perturbed by Gaussian
impurities, the high-energy eigenstates of the system are strongly scarred by
POs of the unperturbed system, and the shape of these scars can be
conveniently tuned via the magnetic field. Furthermore, in the last part of the
work we replace the impurities with a single perturbation, representing a
controllable nanotip,~\cite{nanotip} and use the ``pinning'' property of the
scars to control the scars with the nanotip. Together, these methods of
controlling the strong, perturbation-induced scars could be used to coherently
modulate quantum transport in nanoscale quantum dots.

\section{Model system}

All values and equations below are given in atomic units (a.u.).
The Hamiltonian for a perturbed 2D quantum harmonic oscillator is
\begin{equation}\label{quantum_Hamiltonian}
  H = \frac{1}{2}\big( -\mathrm{i}\nabla + \mathbf{A} \big)^2 + \frac{1}{2} \omega_0^2 r^2 + V_{\text{imp}},
\end{equation}
where $\mathbf{A}$ is the vector potential of the magnetic field. We set the
confinement strength~$\omega_0$ to unity for convenience and assume the
magnetic field~$\mathbf{B}$ is oriented perpendicular to the 2D plane.
We model the perturbation $V_{\textrm{imp}}$ as a sum of Gaussian bumps with
amplitude $M$, that is,
\begin{displaymath}
V_{\text{imp}}(\mathbf{r}) = \sum_{i} M\exp\Bigg[ -\frac{(\mathbf{r} - \mathbf{r}_i)^2}{2\sigma^2}\Bigg].
\end{displaymath}
We focus on the case where the the bumps are positioned randomly with a uniform
mean density of two bumps per unit square. In the energy range considered here,
$E = 50\dots100$, hundreds of bumps exist in the classically allowed region.
The full width at half maximum (FWHM) of the Gaussian bumps
$2\sqrt{2\ln2}\sigma$ is 0.235, which is comparable to the local wavelength of
the eigenstates considered. The amplitude of the bumps is set to $M=4$, which
causes strong PI scarring in this energy regime.

We solve the eigenstates of the Hamiltonian of Eq.~(\ref{quantum_Hamiltonian})
using the \texttt{itp2d} code.~\cite{itp2d} This code utilizes the imaginary
time propagation method, which is particularly suited for 2D problems with
strong perpendicular magnetic fields because of the existence of an exact
factorization of the exponential kinetic energy operator in a magnetic
field.~\cite{itpmagn} The eigenstates and energies
can be compared to the well-known solutions of a unperturbed system. The
unperturbed energies correspond to the Fock-Darwin (FD)
spectrum,~\cite{Fock-Darwin}
\begin{equation}
E_{k, l}^{\text{FD}} = (2k + \vert l \vert + 1)\sqrt{1 + \frac{1}{4}B^2} - \frac{1}{2}lB,
\end{equation}
where $k \in \mathbb{N}$ and $l \in \mathbb{Z}$.
In the limit $B \rightarrow \infty$, the states condensate into Landau levels. The FD spectrum is observed
experimentally in semiconductor quantum dots (see, e.g., Ref.~\onlinecite{Raymond}).


Before considering further the quantum solutions of the Hamiltonian we briefly discuss
the corresponding classical system. The unperturbed 2D harmonic oscillator in a
perpendicular magnetic field is analytically solvable as described in the
Appendix and Ref.~\onlinecite{Kotkin}. The POs
are associated with classical resonances where the oscillation frequencies of the radial
and the angular motion are commensurable. In the following, the notation
($v_{\theta},v_r$) refers to a resonance where the orbit circles the origin
$v_{\theta}$ times in $v_r$ radial oscillations. The resonances occur only at
certain values of the magnetic field given by
\begin{equation}\label{magical_values}
B = \frac{v_r/v_{\theta} - 2}{ \sqrt{v_r/v_{\theta} - 1}}.
\end{equation}

By classical simulations using the \texttt{bill2d} program~\cite{bill2d} we
have confirmed that a perturbation with amplitude~$M=4$ is sufficient to
destroy classical long-term stability in the system. Any remaining
structures in the otherwise chaotic Poincar\'e surface of a section are vanishingly small compared to $\hbar = 1$.

\section{Quantum scars}

\begin{figure}
  \centering
  \includegraphics[width=8.6cm]{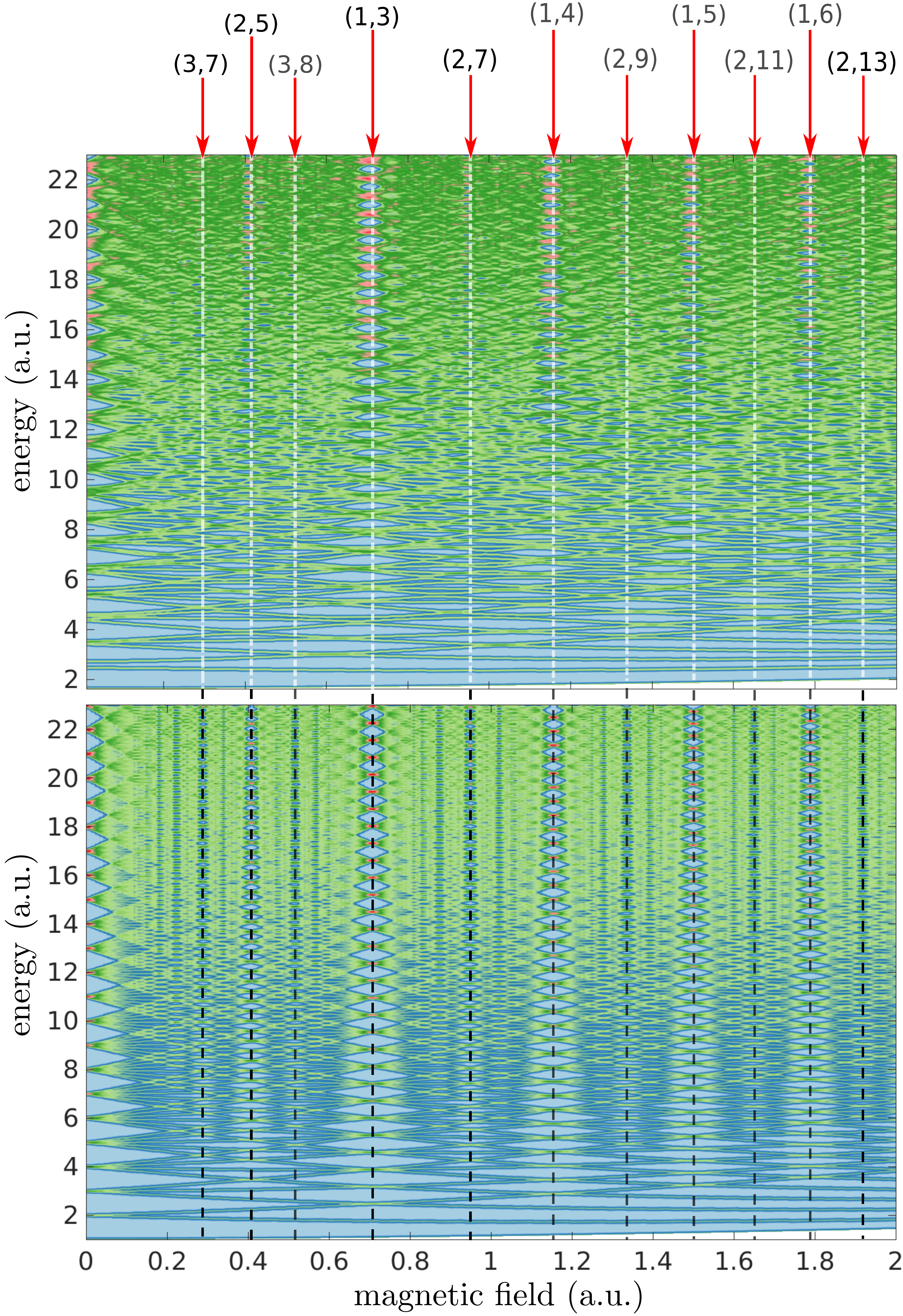}
  \caption{Scaled (arb. units) density of states (DOS) as a function of the
	magnetic field for an unperturbed (lower panel) and perturbed (upper panel)
	two-dimensional harmonic oscillator. The dashed vertical lines indicate the
	resonances ($v_{\theta},v_r$) that correspond to a substantial abundance of
  scarred eigenstates in the perturbed case (see Fig.~\ref{scars}).}
  \label{DOS}
\end{figure}

Next we describe the quantum solutions of the system.
To visualize the spectrum of a few thousand lowest energies as a function of $B$
we show in Fig.~\ref{DOS} the density of states (DOS) computed as a sum of the states
with a Gaussian energy window of $0.001$ a.u. The upper and lower panel correspond to the perturbed (with bumps)
and unperturbed system, respectively. The $B$ values indicating resonances ($v_{\theta},v_r$)
are marked by dashed vertical lines. The upper panel of Fig.~\ref{DOS} shows that the bumps are
sufficiently weak to not completely destroy the FD degeneracies.

\begin{figure}
  \centering
  \includegraphics[width=8.6cm]{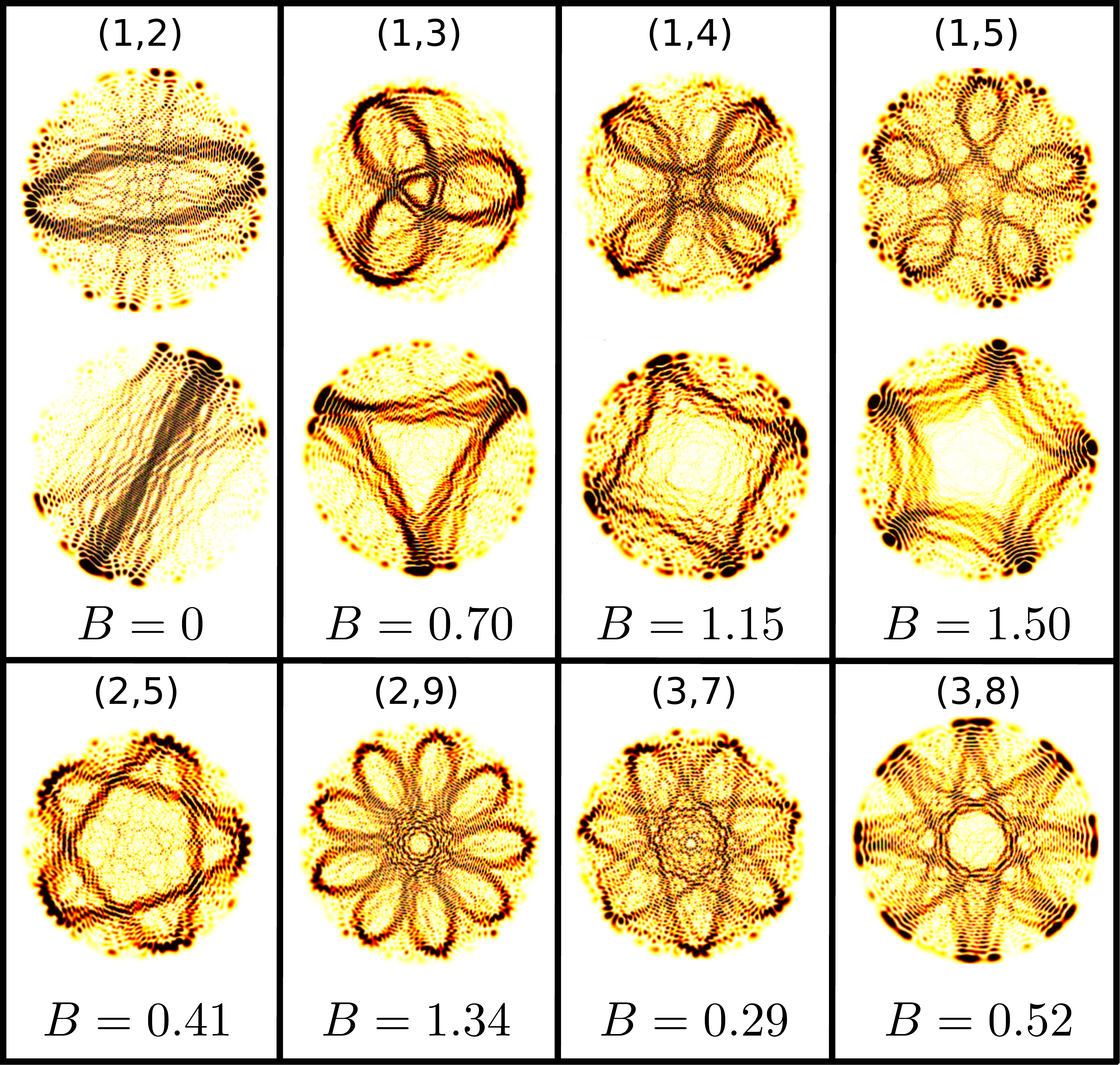}
  \caption{Examples of scars in a perturbed two-dimensional harmonic oscillator
	in a magnetic field. The geometries of the scars can be attributed to the
	classical resonances $(v_{ \theta},v_r)$ and the corresponding periodic
	orbits in the unperturbed system.}
  \label{scars}
\end{figure}

As expected from the theory of PI scarring, the eigenstates of the perturbed
system show clear scars corresponding to the POs of the unperturbed system.
Because of the classical resonance condition \eqref{magical_values} these scars
appear at specific values of the magnetic field where short classical periodic
orbits are possible. Examples of the scars are shown in Fig.~\ref{scars}. The eigenstate number varies between $400\dots3900$. The proportion of strongly scarred states among the eigenstates varies from $10\%$ to $60\%$ at bump
amplitude $M=4$. The proportion and strength of scarred states depend on the
degree of degeneracy in the spectrum: more and stronger scars appear when more
energy levels are (nearly) crossing. The specific shape of the scar for a
chosen rsonance ($v_{\theta},v_r$) depends on the energy. For example, the two examples
shown for the triangular geometry (1,3) in Fig.~\ref{scars} correspond to
orbits with opposite directions, which are not equivalent because the magnetic
field breaks time-reversal symmetry.

\begin{figure}
  \centering
  \includegraphics[width=8.0cm]{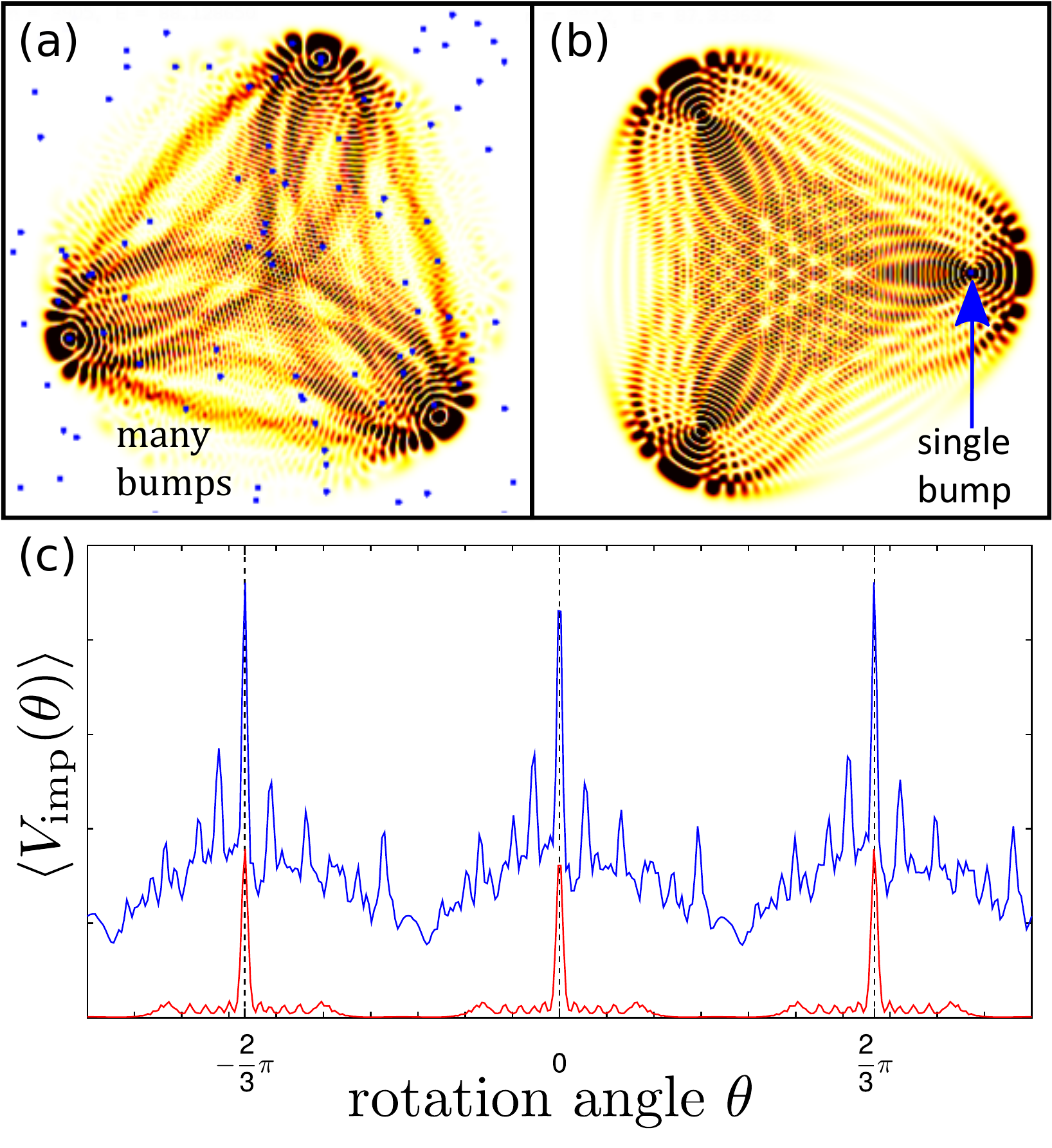}
  \caption{(a)-(b) Examples of scarred eigenstates at $B \approx 0.7$ found in a system
	with several impurities and a single impurity with $M=4$, respectively. The
	locations and FWHMs of the impurities are denoted with blue dots. (c)
	Overlap between the scarred state and the impurity potential, $\langle
	\psi|V_{\text{imp}}(\theta)|\psi \rangle$, as a function of the rotation
	angle with respect to the original orientation $V_{\text{imp}}$. The upper
	(blue) and lower (red) curves correspond to the situations in (a) and (b),
	respectively. The lower curve has been multiplied by a factor of six for
  visibility.}\label{pinning}
\end{figure}

As pointed out in Ref.~\onlinecite{Luukko}, because the scarred states maximize the
perturbation~$V_\text{imp}$ they are oriented to coincide with as many bumps as
possible. This ``pinning'' effect is demonstrated in Fig.~\ref{pinning}(a).
Even a \emph{single} perturbation bump can produce a strong scar that is pinned
to its location, as illustrated in Fig.~\ref{pinning}(b).

As another view on the pinning effect, Fig.~\ref{pinning}(c) shows the overlap
between the scarred state and the impurity potential, $\langle
\psi|V_{\text{imp}}(\theta)|\psi \rangle$, as a function of the rotation angle
with respect to the original orientation $V_{\text{imp}}$. This is shown for
both the multiple-bump (upper curve) and the single-bump (lower curve) cases,
corresponding to situations in Figs.~\ref{pinning}(a) and (b), respectively. The
three distinctive maxima in both cases confirm that the scar is pinned to a location
that maximizes the perturbation.

\begin{figure}
  \centering
  \includegraphics[width=8.0cm]{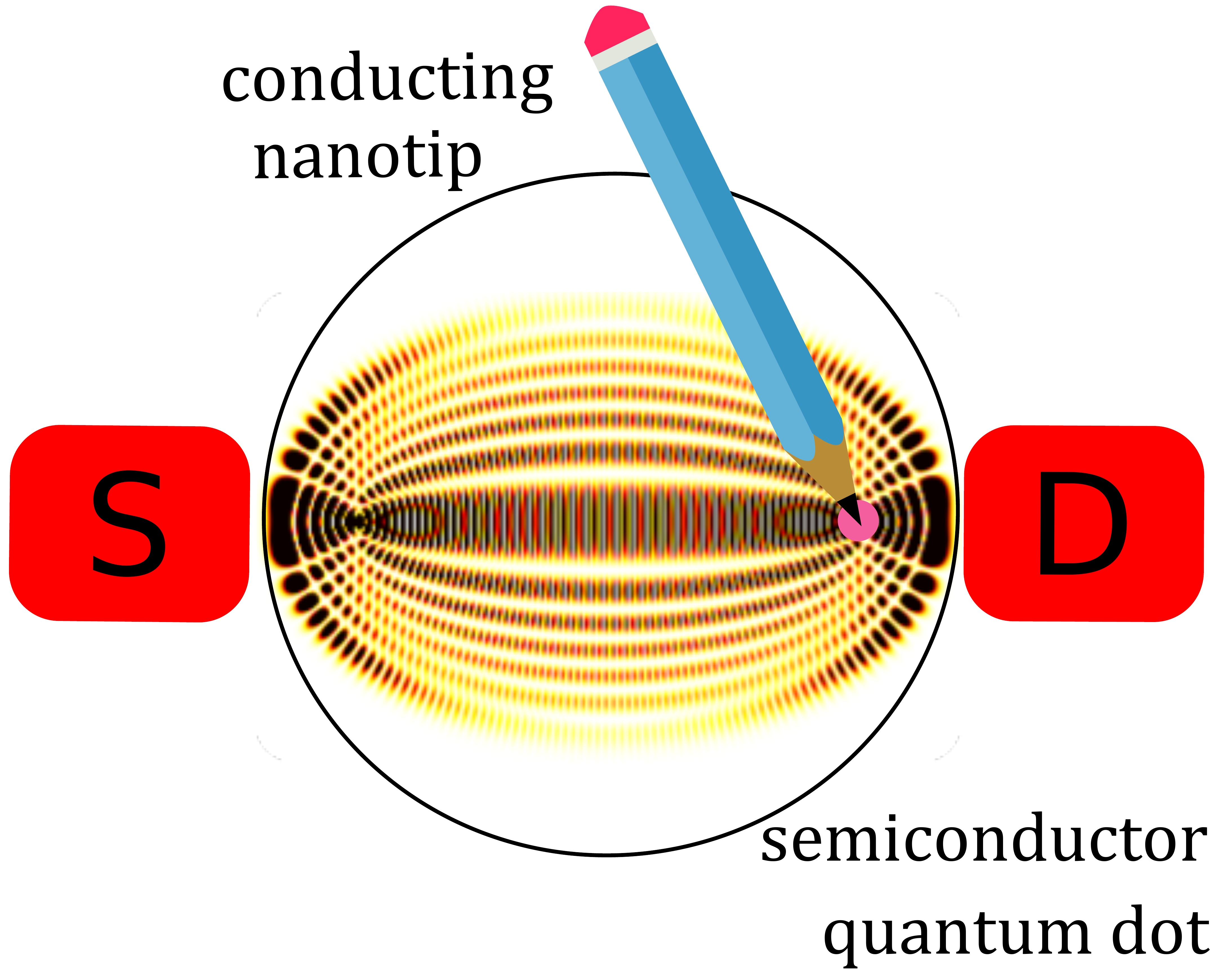}
  \caption{Schematic figure on the utilization of scars in nanostructures. The
	magnetic field (here $B=0$) determines the geometry of the scar (here
	linear), the energy range refines the geometry further (here longitudinal),
	and an external voltage gate generates a local perturbation (here the pink
  dot) that pins the scar, thus determining the orientation.}
  \label{nanotip_figure}
\end{figure}

By creating a single perturbation with, e.g., a conducting
nanotip,~\cite{nanotip} the pinning effect can be used to force the scars to a
specific orientation. Together, the magnetic field and the pinning effect
provide complete external control over the strength, shape, and orientation of
the scars. By selecting which parts of the system are connected by the scar
paths, this provides a method to modulate quantum transport in nanostructures
in a scheme depicted in Fig.~\ref{nanotip_figure}. A successful experiment 
requires that the system is otherwise clean
from external impurities such as migrated ions from the substrate.~\cite{konemann}
In addition, a sufficient energy range in transport is required as the
scars are more abundant at high energies (level number $\gtrsim 500$). In the future, we will study
this effect in more detail using realistic quantum transport calculations.

\section{Summary}

To summarize, we have shown that perturbation-induced quantum scars are found
in a two-dimensional harmonic oscillator exposed to an external magnetic field.
The scars are relatively strong, and their abundance and geometry can be
controlled by tuning the magnetic field. By controlling the orientation of the
scars though their tendency to ``pin'' to the local perturbations, we reach
perfect control of the scarring using external parameters. This scheme, using a
conducting nanotip as the pinning impurity, may open up a path into
``scartronics'', where scars are exploited to coherently control quantum
trasport in nanoscale quantum systems.

\section{Acknowledgments}

We are grateful to Janne Solanp{\"a}{\"a} for useful discussions.
This work was supported by the Academy of Finland.
We also acknowledge CSC -- Finnish IT Center for Science for computational resources.

\appendix

\section{Classical periodic orbits of a two-dimensional harmonic oscillator in a perpendicular and homogeneous magnetic field}

Here we present the solution of the equations of motion for a classical
particle in a harmonic potential with a perpendicular, homogeneous magnetic
field $B$ [see also G. L. Kotkin and V. G. Serbo, \emph{Collection of Problems in Classical Mechanics} (Pergamon Press Ltd., 1971))]. After choosing the vector potential as $A_{\phi} =
\frac{1}{2}Br$, the system is described, in polar coordinates, by the
Lagrangian
\begin{displaymath}
L = \frac{1}{2}(\dot{r}^2 + r^2 \dot{\phi}^2) - \frac{1}{2}\omega_0^2 r^2 +\frac{1}{2}Br^2 \dot{\phi}.
\end{displaymath}
Its constants of motion are the energy $E$ and the angular momentum $p_{\phi}$.

In a rotating frame with $\theta  = \phi + \frac{1}{2} B t$, the Lagrangian has
the form of a simple harmonic oscillator,
\begin{displaymath}
L = \frac{1}{2}(\dot{r}^2 + r^2 \dot{\theta}^2) - \frac{1}{2}\tilde{\omega}^2 r^2,
\end{displaymath}
where the effective frequency is
\begin{displaymath}
\tilde{\omega} = \sqrt{\omega_0^2 + \frac{1}{4}B^2}.
\end{displaymath}
Thus, the system is reduced to a harmonic oscillator without a magnetic field,
the solution of which can be found in standard texts on classical
mechanics [e.g.,  H. Goldstein, \emph{Classical Mechanics}, 2nd ed. (Addison-
Wesley, 1980)]. When the initial conditions are chosen as $\phi(0)
= 0$ and $r(0) = a$, the solution in the original coordinates $(r, \phi)$ is
\begin{displaymath}
r^2 = a^2\cos^2(\tilde{\omega}t) + b^2\sin^2(\tilde{\omega}t)
\end{displaymath}
and
\begin{displaymath}
\phi(t) = -\frac{1}{2}Bt + \arctan \Big[ \frac{a}{b} \tan(\tilde{\omega}t) \Big],
\end{displaymath}
where the branches of the arctangent are chosen such that the angle $\phi$ is a
continuous function of time $t$. The maximum and minimum radii, $a$ and $b$,
are determined by the angular momentum
\(
p_{\phi} = \tilde{\omega}ab
\)
and the energy
\begin{displaymath}
E = \frac{1}{2} \tilde{\omega}^2 (a^2 + b^2) - \frac{1}{2}p_{\phi}B.
\end{displaymath}

The solution shows that the period of radial oscillations $T = \pi/
\tilde{\omega}$ is independent of $E$ and $p_{\phi}$.
The angle with which the radius vector turns during a period is
\begin{equation}\label{Delta_phi}
\Delta \phi = \pi\Big[ \xi(p_{\phi}) - \frac{B}{2 \tilde{\omega}} \Big],
\end{equation}
where the step function $\xi(x)$ is
\begin{displaymath}
\xi(x) =
\left\{ \begin{array}{ll}
-1 & \textrm{if $x < 0$}\\
\phantom{-} 0 & \textrm{if $x = 0$}\\
\phantom{-}1 & \textrm{if $x > 0$.}
\end{array} \right.
\end{displaymath}
Note that $\Delta \phi$ is also independent of $E$.

The orbit is periodic exactly when $\Delta \phi$ is a rational multiple of
$2\pi$. Other orbits are quasiperiodic. When $\Delta \phi = 2\pi
v_{\theta}/v_r$, the particle returns to its initial state after rotating
$v_{\theta}$ times around the origin and oscillating $v_r$ times between the
radial turning points. Thus, the classical oscillation frequencies are said to
be in $v_{\theta}:v_r$ resonance for this particular PO.

By solving Eq.~(\ref{Delta_phi}) for the magnetic field, we can determine which
value of~$B$ is required for a $v_{\theta}:v_r$ resonance. This gives, after
elementary algebra, the resonance condition
\begin{equation*}
B = \frac{v_r/v_{\theta} - 2}{ \sqrt{v_r/v_{\theta} - 1}}.
\end{equation*}


\begin{thebibliography}{99}

\bibitem{Heller} E. J. Heller, \href{https://doi.org/10.1103/PhysRevLett.53.1515}{Phys. Rev. Lett. $\boldsymbol{53}$, 1515-1518 (1984)}.
\bibitem{Kaplan} L. Kaplan, \href{http://stacks.iop.org/0951-7715/12/i=2/a=009}{Nonlinearity $\boldsymbol{12}$, R1 (1999)}.

\bibitem{Luukko} P. J. J. Luukko, B. Drury, A. Klales, L. Kaplan, E. J. Heller, and E. R{\"a}s{\"a}nen, \href{http://dx.doi.org/10.1038/srep37656}{Sci. Rep. $\boldsymbol{6}$, 37656 (2016)}.

\bibitem{reimannmanninen}  S. M. Reimann and M. Manninen, \href{https://doi.org/10.1103/RevModPhys.74.1283}{Rev. Mod. Phys. {\bf 74}, 1283 (2002)}.
\bibitem{rasanen} E. R\"as\"anen, H. Saarikoski, A. Harju, M. Ciorga, and A. S. Sachrajda, \href{https://doi.org/10.1103/PhysRevB.77.041302}{Phys. Rev. B {\bf 77}, 041302(R) (2008)}.
\bibitem{rogge} M. C. Rogge, E. R\"as\"anen, and R. J. Haug, \href{https://doi.org/10.1103/PhysRevLett.105.046802}{Phys. Rev. Lett. {\bf 105}, 046802 (2010)}.
\bibitem{konemann} E. R\"as\"anen, J. K\"onemann, R. J. Haug, M. J. Puska, and R. M. Nieminen, \href{https://doi.org/10.1103/PhysRevB.87.241303}{Phys. Rev. B {\bf 70}, 115308 (2004)}.

\bibitem{nanotip}  A. C. Bleszynski, F. A. Zwanenburg, R. M. Westervelt,
A. L. Roest, E. P. A. M. Bakkers, and L. P. Kouwenhoven,
\href{https://doi.org/10.1021/nl0621037}{Nano Lett. $\boldsymbol{7}$, 2559 (2007)}; E. E. Boyd, K. Storm,
L. Samuelson, and R. M. Westervelt, \href{http://stacks.iop.org/0957-4484/22/i=18/a=185201}{Nanotechnology
$\boldsymbol{22}$, 185201 (2011)}; T. Blasi, M. F. Borunda, E. R{\"a}s{\"a}nen,
and E. J. Heller, \href{https://doi.org/10.1103/PhysRevB.87.241303}{Phys. Rev. B $\boldsymbol{87}$, 241303 (2013)}.

\bibitem{itp2d} P. J. J. Luukko and E. R\"as\"anen, \href{http://dx.doi.org/10.1016/j.cpc.2012.09.029}{Comp. Phys. Comm. $\boldsymbol{184}$, 769 (2013)}.
\bibitem{itpmagn} M. Aichinger, S. A. Chin, and E. Krotscheck, \href{https://doi.org/10.1016/j.cpc.2005.05.006}{Comput.~Phys.~Commun. $\boldsymbol{171}$, 197 (2005)}.
\bibitem{Fock-Darwin} V. Fock, \href{https://doi.org/10.1007/BF01390750}{Z. Phys. $\boldsymbol{47}$, 446 (1928)}; C. G. Darwin, \href{}{Proc. Cambr. Philos. Soc. $\boldsymbol{27}$, 86 (1930)}.
\bibitem{Raymond} S. Raymond, S. Studenikin, A. Sachrajda, Z. Wasilewski, S. J. Cheng, W. Sheng, P. Hawrylak, A. Babinski,
M. Potemski, G. Ortner, and M. Bayer, \href{https://doi.org/10.1103/PhysRevLett.92.187402}{Phys. Rev. Lett. $\boldsymbol{92}$, 187402 (2004)}.

\bibitem{Kotkin} G. L. Kotkin and V. G. Serbo, \emph{Collection of Problems in Classical Mechanics} (Pergamon Press Ltd., 1971).
\bibitem{bill2d} J. Solanp{\"a}{\"a}, P. J. J. Luukko, and E. R{\"a}s{\"a}nen, \href{http://dx.doi.org/10.1016/j.cpc.2015.10.006}{Comp. Phys. Comm. $\boldsymbol{199}$, 133 (2016)}.
\end{thebibliography}
\end{document}